\documentclass[12pt,preprint]{aastex}

\shorttitle{New gas cell for near infrared calibration}
\shortauthors{Valdivielso et al.}

\begin{document}


\title{A New  Gas Cell for High-Precision Doppler Measurements in the Near-Infrared}


\author{L. Valdivielso}%\altaffilmark{1}}
\affil{Instituto de Astrof\'isica de Canarias, C/V\'ia L\'actea, s/n, 38205, La Laguna, Tenerife, Spain}
\email{lval@iac.es}
\author{P. Esparza}%\altaffilmark{2}}
\affil{ Dpto. de Qu\'imica Inorg\'anica de la Universidad de la Laguna, C/ Francisco S\'anchez, s/n, 38204, La Laguna, Tenerife, Spain}
\author{ E. L. Mart\'in\altaffilmark{1}}
\affil{ CSIC-INTA Centro de Astrobiolog\'ia, Ctra. Ajalvir km 4, 28850, Torrej\'on de Ardoz, Madrid, Spain }

\and

\author{D. Maukonen  and  R. E. Peale}
\affil{University of Central Florida, Physics Department, Orlando, Fl 32816, USA}


\altaffiltext{1}{University of Central Florida, Physics Department, Orlando, Fl 32816, USA}


\begin{abstract}
High-resolution spectroscopy in the near-infrared could become the leading method for discovering extra-solar planets around very low-mass stars and brown dwarfs. To help to achieve an accuracy of $\sim$\,m/s, we are developing a gas cell which consists of a mixture of gases whose absorption spectral lines span all over the near-infrared region. We present the most promising mixture, made of acetylene, nitrous oxide, ammonia, chloromethans and hydrocarbons. The mixture is contained in a small size 13\,cm long gas cell and covers most of the H and K-bands. It also shows small absorptions in the J-band but they are few and not sharp enough for near infrared wavelength calibration. We describe the working method and experiments and compare our results with the state of the art for near infrared gas cells.  
\end{abstract}





\section{Introduction}

In the last few years, a new generation of near infrared (NIR)
spectrographs with high spectral resolution and radial velocity
accuracy ($\sim$ m/s) for exoplanet detection is under development
\citep{2004SPIE.5492.1274O,2005AN....326.1015M,2008PASP..120..887R}.
For this purpose, a high precision calibration system with a stable
and very high number of lines spanning all over the instrumental
wavelength range is mandatory.  Available Thorium-Argon (Th-Ar)
emission lamps \citep{2007A&A...468.1115L} provide a good coverage in
the optical. However, there is a lack of such lines in the near
infrared, and those available are relatively faint and unstable
(primary Ar, as they are the dominant brightness lines in the $1-1.8
\mu$m region; Wahling et al. 2002). Laser frequency combs have the
advantage of generating series of equally spaced very narrow lines
\citep{2007MNRAS.380..839M,2008Sci...321.1335S} but stability and
repeatability in long timescales has not been achieved yet. \\ In the
optical regime, iodine gas cell has proved to be a very good method
for simultaneous calibration of echelle spectra
\citep{1996PASP..108..500B}. This method has several advantages
compared to others. The cell is located along the stellar beam in
front of the spectrograph. Thus, the stellar spectrum is superimposed
to the absorption spectrum for simultaneous calibration so that
unstabilities and differences in illumination of the spectrograph can
be measured and modelled to remove these effects. It is also very
cheap (specially compared to laser combs) and of easy implementation
and maintenance but since there are no pure gases showing a wide
spectral domain in the near infrared, as iodine in the optical, the
main problem is to get a suitable gas cell with strong and well
distributed absorption lines, stable with time and
temperature.\\ \cite{2009ApJ...692.1590M} have considered series of
commercially available absorption cells of H$^{13}$C$^{14}$N,
$^{12}$C${_2}$H${_2}$, $^{12}$CO and $^{13}$CO for the H-band and
\cite{2008SPIE.7014E.126D} have simulated a mixture of HCl, HBr and HI
gas cell for the GIANO spectrograph in the 0.9--2.5\,$\micron$ range
using the HITRAN database \citep{2005JQSRT..96..139R}.  Also a couple
of instruments are already using absorption gas cells that cover small
parts of the near infrared spectrum. A N$_2$O gas cell
\citep{2007ASPC..364..461K} has been developed for the CRIRES
spectrograph.  

Also, recently, an ammonia gas cell has been used by
\cite{2009arXiv0911.3148B} in the K-band, and they have proven that
such a cell can achieve precisions of $\sim$ 5 m/s over long
timescales with CRIRES, which proves the feasibility of this technique
also in the near infrared. \cite{2003A&A...403.1077K} reported a
radial velocity precision of 2.65\,m/s for Barnard's star (spectral
type M4) and they found that the radial velocity is limited by stellar
noise (activity and convection). \\

Over the last years we have been working on a
gas cell for simultaneous calibration of spectra from the Y, J, H and
K-bands. In this paper we present the experimental aproach and the
promising results obtained so far. The paper is organized as
follows. In section~\ref{2}, we describe the gas cell properties and
laboratory experiments to construct the gas mixtures. In
section~\ref{3} we describe the laboratory measurements and present
the results obtained. We conclude in section~\ref{4} with a
discussion.

\section{Gas cell properties and filling process}\label{2}
The gas samples were contained in an International Crystal Lab (ICL) model G-2 gas cell at atmospheric pressure. We have used 10cm and 13cm length gas cells with a window size of 38mm diameter and 6mm thickness of infrasil material designed to permit infrared analysis of low volume gases. The windows are glued to the main body of the cell so that gases remain insulated from the outside environment, which is important in order to avoid chemical contamination.\\ 
Samples were introduced into the cell via a vacuum line system built in our laboratory. The gases are mixed in a controlled high vacuum chamber with liquid nitrogen trap and with manual manometric control using a mercury column. After reaching vacuum inside the cell, different partial pressures of each gas are injected. The mixture is completed with Argon gas, which is chemically inert and shows no absorption lines, up to atmospheric pressure for stability.
Initially we have used single gases in order to study potentially good candidates for the mixtures. For this purpose we filled the cells with 50\% gas $+$ 50\% Argon.

\section{Results}\label{3}
For an accurate calibration we search for a gas mixture with a forest of lines well distributed in the spectral interval between 0.9 and 2.5 $\mu$m. The working method consists in obtaining measurements from 800 to 2500 nm of the selected individual gases in order to check whether they show any absorption band useful for a mixture, and those potentially good were used to make a gas mixture and study its stability with time and temperature.\\
For laboratory measurements we have used the Cary 5 spectrophotometer placed on the Instituto de Astrof\'isica de Canarias optical laboratory. This instrument is helpful for absorption spectroscopy in the 175--3300nm range with a PbS detector cooled to 0$\degr$C for photometric noise reduction in the near-infrared. The measurements were performed at a mean laboratory temperature of 24\,$\degr$C. We have used Ar gas cells as reference in order to apply a baseline correction to the sample scan that is collected from the instrument as transmission values. Since we are interested in finding absorption bands, the measurements have been done with a fixed spectral bandwidth of 2\,nm (FWHM) sample at 1\,nm. This provides a high light throughput and good signal to noise performance, and is enough for sampling the absorption bands we are searching for despite of the low resolution of the scan (R $\equiv \lambda/\Delta\lambda =$ 1250 at 2.5\,$\mu$m).\\

\begin{figure}
\includegraphics[scale=0.8]{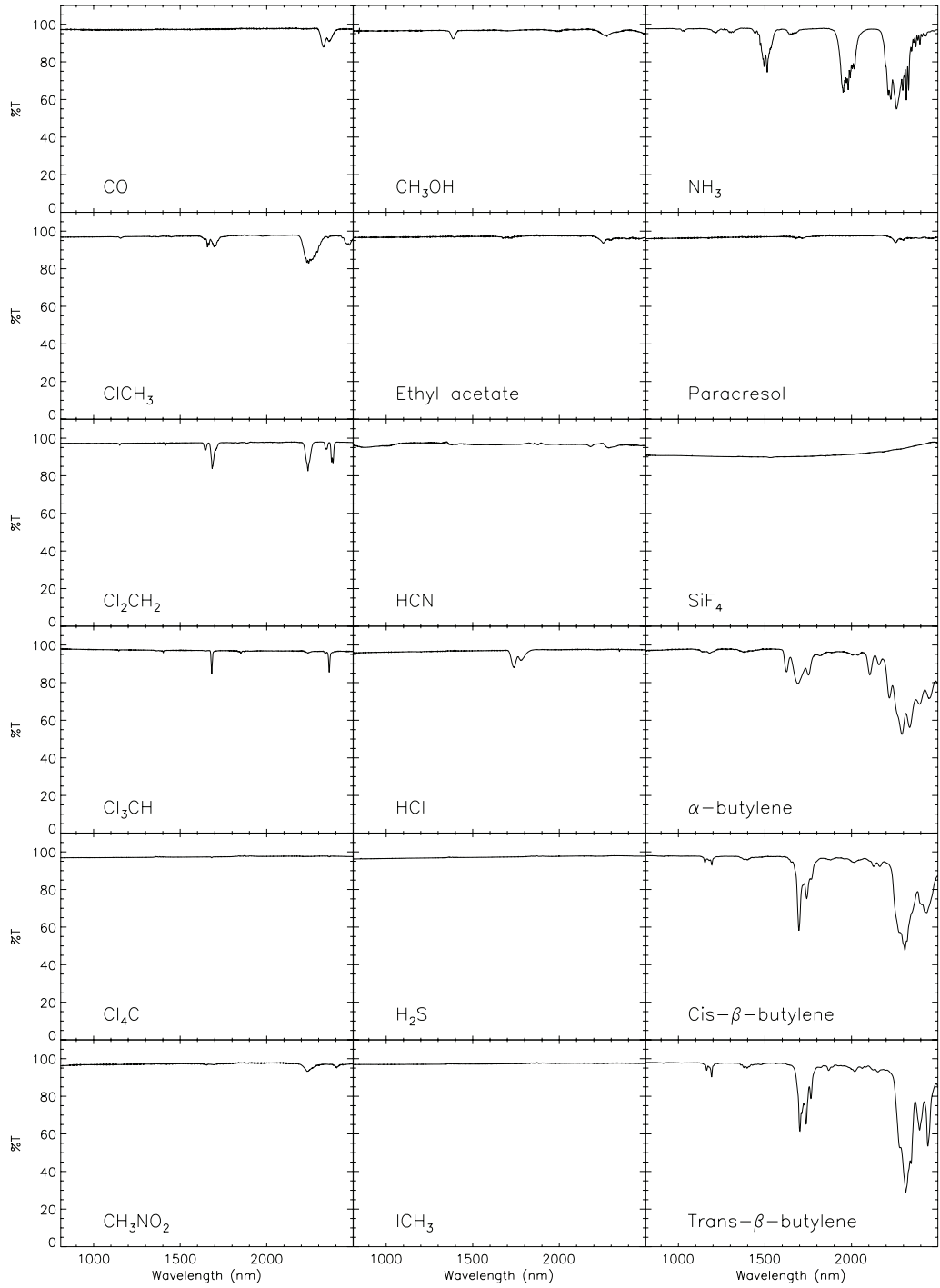}
\caption{Low resolution near infrared spectra of 18 individual gases. The cells were filled with 50\% gas $+$ 50\% Argon. 
        }\label{fig1}
\end{figure}

\begin{figure}
\includegraphics[scale=1.0]{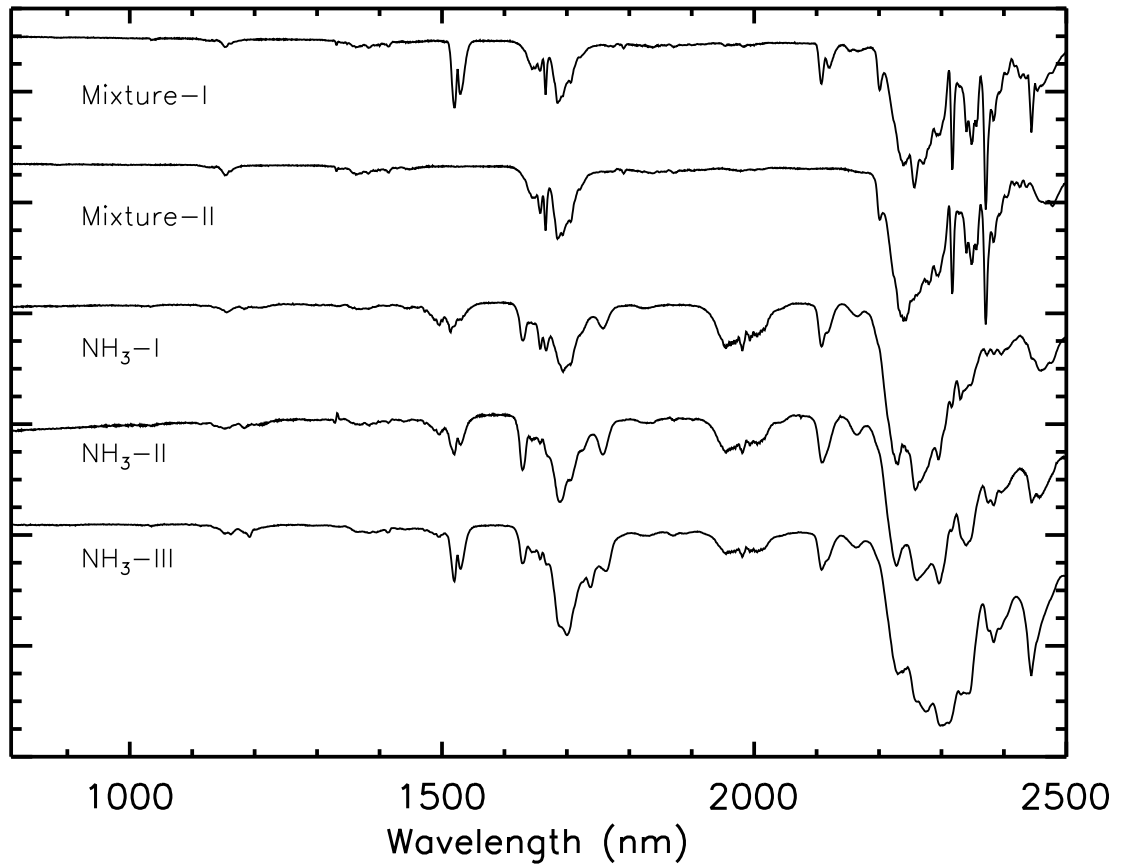}
\caption{Low resolution near infrared spectra of five gas mixtures obtained at IAC optical laboratory.
        }\label{fig2}
\end{figure}

A first report was presented in \cite{2005AN....326.1015M} describing the first experiments with a gas cell filled with N$_{2}$O, H$_{2}$C$_{2}$ and CH$_{4}$. We checked that the cell  was stable over a timescale of one year at least. We have investigated 18 new single gases. In Figure \ref{fig1} we show the data obtained for each gas. They are listed in Table \ref{tab1}  and we note if they show clear absorption or not.\\ 
Seven of them show negligible absorption, or do not show any band or line so they were discarded for the gas mixtures. We have chosen high purity gases commercially available with deeper absorptions and wider spectral coverage, adding and/or replacing individual gases with new ones, including acetylene, nitrous oxide, hydrocarbons and chloromethans, using different partial pressures. All of them are safe to use at this small concentrations, are not corrosive and are gaseous at room temperature; this makes them suitable  for regular ground-based observations.\\
 In total we have worked with five new different mixtures (see  Figure \ref{fig2}). We have listed the composition in Table \ref{tab2}. Due to the number of gas cells available, only three have been measured twice for stability study. The first two new mixtures (namely Mixture-I, II), included nitrous oxide, acetylene, methane and/or chloromethans; only Mixture-I was measured on two occasions three months apart. We found Mixture-I to be stable, with differences on the measurements lower than 1\% on band absorption intensity and 0.5\,nm on line position. Based on this mixture, three additional mixtures introducing ammonia and hydrocarbons were produced, namely NH$_{3}$-I, II and III. The mixture NH$_{3}$-II was measured eight months apart and remained stable with time and temperature, but we have discarded methane because of the atmospheric absorption. \\

We have carried out a characterization of the NH$_{3}$-III gas cell by means of an Infrared Fourier Transform (FTIR) Spectrometer (Bomen DA8), located at the University of Central Florida (USA). The instrument makes use of a InSb ( Indium Antimonide) detector, a quartz-halogen lamp as source and a quartz beam-splitter. The spectrum was obtained under vacuum ($<$ 5mTorr) at 23$\degr$C with a resolution of 0.1 cm$^{-1}$ (R = 40000 at 2.5 $\mu$m). \\
The cell was measured with Cary 5 and on two separate occasions with FTIR three and eight months apart. Figure \ref{3} compares a high resolution measurement with the low resolution measurement, an ammonia gas cell (50\% NH$_{3}$ + 50\% Argon), a M9 brown dwarf model (R $\sim 65000$) and a telluric spectra in absorption (from top to bottom). This gas cell consists on a mixture of  N$_{2}$O, H$_{2}$C$_{2}$, ClCH$_{3}$, Cl$_{2}$CH$_{2}$, NH$_{3}$, $\alpha$-Butylene and  Trans-$\beta$-butylene. \\

There are two main bands covering a significant fraction of the H-band
(red vertical lines) which are not heavily affected by telluric
absorptions and thus make them object of interest for accurate
velocity measurements. The first one is clearly seen as a forest from
1.47--1.54 $\mu$m, mainly due to acetylene absorption (53 lines
resolved at 0.1 cm$^{-1}$ between 1.51--1.54\,$\mu$m; R$\sim 66000$)
but also ammonia. Acetylene was one of the first candidates and was
already used in the first mixture presented in 2005. It exhibits
absorptions not only in the 1.51--1.54\,$\mu$m wavelength range but
also in the 2.4--2.5\,$\mu$m although this region of the near infrared
is severely affected by the atmospheric contamination.  The second
band covering the H-band (1.62--1.78 $\mu$m) is a combination of
chloromethans and hydrocarbons and appears mostly pressure
broadened. The same phenomena appears in the K-band. There is a clear
forest of lines between 1.8--2.09 $\mu$m which, by comparison with the
third spectra from the top, can be clearly assessed to ammonia
absorption, and another pressure broadened band in the 2.1-2.5 $\mu$m
wavelength range, with absorptions of all of the individual gases,
specially hydrocarbons. The combination of the presence of suitable
cell in the H-band and the simulations that show that this band
contains suitable Doppler information in M dwarfs
(\citeauthor{2010ApJ...710..432R} \citeyear{2010ApJ...710..432R};
Rodler et al. 2010, in preparation) indicates that this near-infrared
spectral window offers the best possibility for high-precision radial
velocity work in M dwarfs.\\ There is a lack of lines in the
1.54--1.62 $\mu$m wavelength range. \citet{2009ApJ...692.1590M}
explored series of gas cells for the H-band covering the 1.51--1.63
$\mu$m wavelength range also including C${_2}$H${_2}$, CO and HCN. In
Fig \ref{fig1} we can see our laboratory absorption spectra of CO and
HCN using 50\% partial pressures on a 10\,cm gas cell. CO shows clear
absorption in the K-band but neither HCN neither CO show absorption in
the H-band with this concentration for such an small gas cell. One of
our objectives is to find a compact gas cell with easy implementation
useful for the new generation near infrared spectrographs, optimising
the space with one single manageable and of low maintentance gas
cell. The absorption gas cell proposed for the GIANO instrument
\citep{2006SPIE.6269E..41O} consists of a mixture of HCl, HBr and HI
\citep{2008SPIE.7014E.126D} on a 0.5\,m gas cell. Hydrogen Iodide
shows line absorptions between 1.53--1.6 $\mu$m which would fill the
lack of lines we have in the H-band but again, obtaining deep
absorption lines would require filling the cell with high
concentrations of this gas and a much more controlled
environment.\\ As we have mentioned previously, an ammonia gas cell
has recently been successfully used on CRIRES for high resolution
measurements \citep {2009arXiv0911.3148B}. They make use of a 17\,cm
gas cell filled with 50 mb ammonia at 15$\degr$\,C which is almost 1/3
of our partial pressure. Ammonia exhibits a high number of lines in
the K-band and despite of the telluric absorption lines of the region
used for observations (grey vertical lines), they have demonstrated
the feasibility of obtaining few m/s radial velocity accuracy using
such a gas cell. We expect to resolve the individual lines using lower
partial pressures in our gas cell with the proper concentration in
order to get deep absorptions without line broadening and blending
covering a wider wavelength range (1.9--2.5\,$\mu$m) with the gas
mixture we are working with. \\ Unfortunately, there are few
absorptions in the Y and J-band. In the range 1.15--1.22 $\mu$m there
is a small contribution of chloromethans and hydrocarbons but deepness
and density of lines is not enough for high accuracy radial velocity
measurements. This is still a problem on the use of gas cells in the
near infrared. There are few suitable gases known with absorptions in
the Y and J-bands. Chloromethans, acetylene and hydrocarbons used here
show small contributions. Also the absorption gas cell proposed by
\citet{2008SPIE.7014E.126D} covers the H and K-bands with $\sim$\,200
lines but again there is a lack for Y and J-bands calibration with few
absorptions. Obtaining deeper absorptions would imply to increase the
length of the gas cell, but would also imply a deeper absorption for
the H and K-bands.\\

\section{Discussion}\label{4}
We have presented new results of gas mixtures for wavelength calibration echelle spectrographs in the near infrared. We have worked on different gas cells including several new gases. The working method and the properties of the gas cells have been described. We have obtained several mixtures and we have presented a compact and manageable gas cell which covers the widest wavelength range to date in the H and K-bands, with a potentially high number of lines than for currently available gas cells, stable in time scales of months under atmospheric temperature conditions of our laboratory, and which can be useful for high precision radial velocity measurements. We work on the improvement of the gas cell using different partial pressures of the individual gases in order to solve the pressure broadening of some absorption bands. Some of these gases have been recently tested with real observations and obtained promising results with few m/s accuracy. Such gas cells can be of interest for several new generation high resolution near infrared spectrographs under development (NAHUAL, GIANO, PRVS, SPIROU, CARMENES).\\

\acknowledgments
This work has been supported by the Spanish Ministerio de Eduaci\'on y Ciencia through grant AYA2004-08271-C01. Work on writing this paper was developed while LV was a visitor at the Centro de Estudios de F\'isica del Cosmos de Arag\'on, whose hospitality is acknowledged gratefully.


%% in parentheses, after the keyword, but they will not be verified.

\begin{table}
\begin{center}
\caption{List of individual gases.\label{tab1}}
\begin{tabular}{lclclc}
\tableline\tableline
Gas & Absorption & Gas & Absorption  & Gas & Absorption \\
 &  band &  &  band &  &  band\\
\tableline
 CO		&yes&	CH$_{3}$OH&yes&NH$_{3}$	&yes\\
ClCH$_{3}$	   &yes&Ethyl acetate       &negligible&Paracresol          &negligible\\
Cl$_{2}$CH$_{2}$    &yes&	HCN&yes&	SiF$_{4}$&none\\
Cl$_{3}$CH          &yes&	HCl		&yes&	$\alpha$-butylene&yes\\ 
Cl$_{4}$C           &none& H$_{2}$S		&none&cis-$\beta$-butylene&yes  \\
CH$_{3}$NO$_{2}$     &negligible&ICH$_{3}$           &none&	trans-$\beta$-butylene&yes\\
\tableline
\end{tabular}
\end{center}
\end{table}

\begin{table}
\begin{center}
\caption{\label{tab2} Composition of five new gas mixtures. Low resolution spectra obtained in the laboratory are shown in Figure \ref{fig2}.}
\begin{tabular}{ll}
\tableline\tableline
Name & Composition \\
\tableline
Mixture-I &    CH$_{4}$+ ClCH$_{3}$ +  Cl$_{2}$CH$_{2}$ + NO$_{2}$ + C$_{2}$H$_{2}$ \\     
Mixture-I &    CH$_{4}$+ ClCH$_{3}$ +  Cl$_{2}$CH$_{2}$ + Cl$_{3}$CH  \\ 	    
NH3-I &         ClCH$_{3}$ + Cl$_{2}$CH$_{2}$ + NO$_{2}$ + C$_{2}$H$_{2}$+ NH$_{3}$ + $\alpha$-butylene \\      	    
NH3-II &        ClCH$_{3}$ + Cl$_{2}$CH$_{2}$ + NO$_{2}$ + C$_{2}$H$_{2}$+ NH$_{3}$ + $\alpha$-butylene +  CH$_{4}$ \\     
NH3-III &       ClCH$_{3}$ + Cl$_{2}$CH$_{2}$ + NO$_{2}$ + C$_{2}$H$_{2}$+ NH$_{3}$ + $\alpha$-butylene +  trans-$\beta$-butylene\\
\tableline
\end{tabular}
\end{center}
\end{table}

\begin{figure}
\includegraphics[scale=1.0]{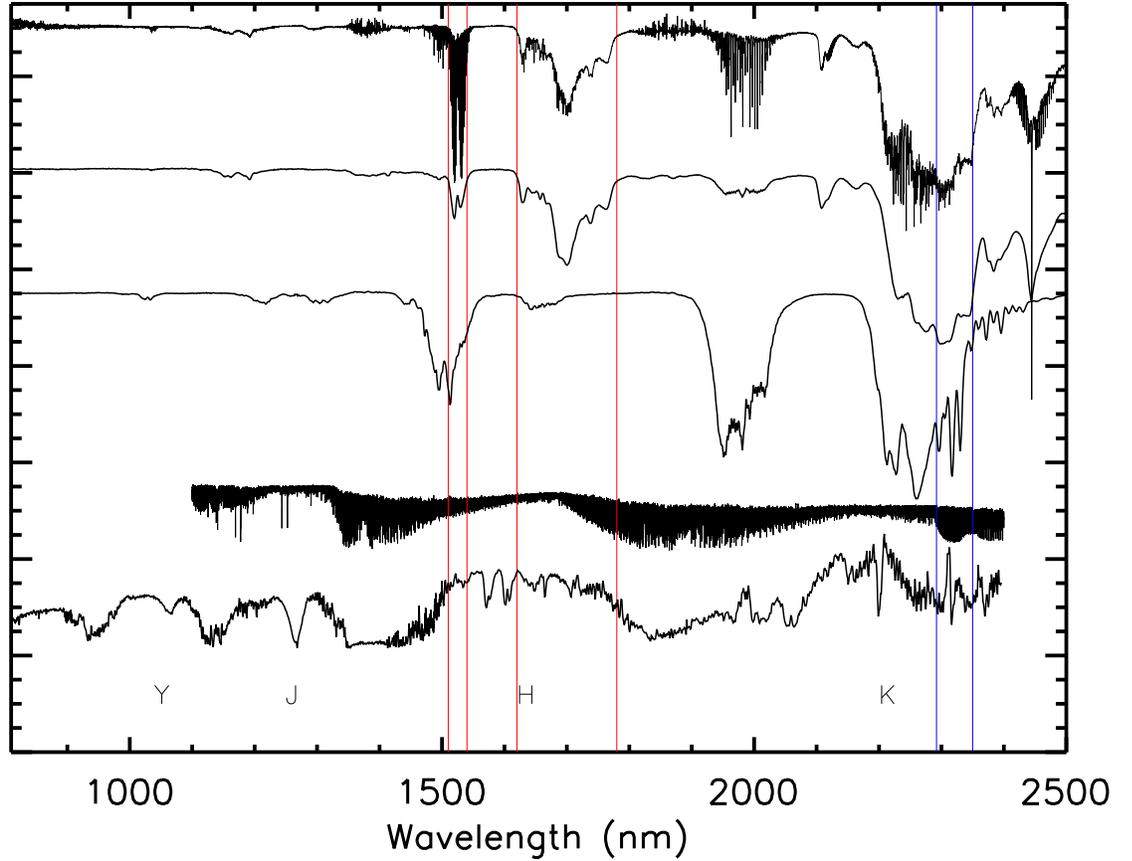}
\caption{Near infrared spectra of  NH$_{3}$-III gas cell obtained with FTIR, Cary 5, an ammonia gas cell (50\% NH$_{3}$ + 50\% Argon), a M9 brown dwarf model (R $\sim$ 65000) and a telluric absorption spectrum from top to bottom. Red vertical lines show the two main absorptions in the H-band. Those regions are examples of wavelength ranges that could be of interest for radial velocity work with CRIRES using our gas cell. Blue vertical lines show the window observed by \cite{2009arXiv0911.3148B}. The standard windows Y, J, H and K-bands are indicated.}\label{fig3}
\end{figure}
\end{document}